\documentclass[fleqn,usenatbib]{mnras}

\usepackage{amsmath}	
\usepackage{amssymb}	

\usepackage{mathptmx}
\usepackage{txfonts}

\usepackage[T1]{fontenc}
\usepackage{ae,aecompl}

\usepackage{graphicx}	

\newcommand{\hete}{HETE J1900.1--2455}

\newcommand{\xte}{\textsl{RXTE}}

\def\Mns{M_{\rm NS}}
\def\Rns{R_{\rm NS}}
\def\Rbb{R_{\rm bb}}
\def\fc{f_{\rm c}}

\def\Msun{{\rm M}_{\odot}}

\def\ergcms{\mathrm{erg\,cm^{-2}\,s^{-1}}}

\def\gs{{\rm g\,s^{-1}}}

\def\gcm2{{\rm g\,cm^{-2}}}

\title[Detection of burning ashes from X-ray bursts]{Detection of burning ashes from thermonuclear X-ray bursts} 

\author[J.~J.~E. Kajava et al.]{
J.~J.~E. Kajava,$^{1}$
J. N\"{a}ttil\"{a},$^{2,3}$
J. Poutanen,$^{2,3}$
A. Cumming,$^{4}$
V. Suleimanov$^{5,6}$\newauthor and 
E. Kuulkers$^{1}$\\
$^{1}$European Space Astronomy Centre (ESA/ESAC), Science Operations Department, 28691 Villanueva de la Ca\~{n}ada, Madrid, Spain\\
$^{2}$Tuorla Observatory, University of Turku, V\"{a}is\"{a}l\"{a}ntie 20, FIN-21500 Piikki\"{o}, Finland\\
$^{3}$Nordita, KTH Royal Institute of Technology and Stockholm University, Roslagstullsbacken 23, SE-10691 Stockholm, Sweden\\
$^{4}$Department of Physics and McGill Space Institute, McGill University, 3600 rue University, Montreal, QC H3A2T8, Canada\\
$^{5}$Institut f\"ur Astronomie und Astrophysik, Kepler Center for Astro and Particle Physics, Universit\"at T\"ubingen, Sand 1, 72076 T\"ubingen, Germany\\
$^{6}$Kazan (Volga region) Federal University,  Kremlevskaya str. 18, Kazan 420008, Russia
}

\date{Accepted 2016 August 23. Received 2016 August 23; in original form 2016 June 15}

\pubyear{2016}

\begin{document}
\label{firstpage}
\pagerange{\pageref{firstpage}--\pageref{lastpage}}
\maketitle

\begin{abstract}
When neutron stars (NS) accrete gas from low-mass binary companions, explosive nuclear burning reactions in the NS envelope fuse hydrogen and helium into heavier elements.
The resulting thermonuclear (type-I) X-ray bursts produce energy spectra that are fit well with black bodies, but a significant number of burst observations show deviations from Planck spectra.
Here we present our analysis of \xte/PCA observations of X-ray bursts from the NS low-mass X-ray binary \hete.
We have discovered that the non-Planckian spectra are caused by photo-ionization edges.
The anti-correlation between the strength of the edges and the colour temperature suggests that the edges are produced by the nuclear burning ashes that have been transported upwards by convection and become exposed at the photosphere.
The atmosphere model fits show that occasionally the photosphere can consist entirely of metals, and that the peculiar changes in black body temperature and radius can be attributed to the emergence and disappearance of metals in the photosphere.
As the metals are detected already in the Eddington-limited phase, it is possible that a radiatively driven wind ejects some of the burning ashes into the interstellar space.
\end{abstract}




\section{Introduction}

The discovery of thermonuclear (type-I) X-ray bursts more than 40 years ago \citep{GGS76} opened a new window to study neutron stars (NS).
The fuel that powers an X-ray burst comes from a low-mass companion star, from which the NS accretes gas through an accretion disc.
For specific ranges of the mass accretion rate (e.g., \citealt{FHM81}), unstable nuclear burning can occur producing elements up to tellurium \citep{SAB01}.
The energy released in the burning escapes as thermal X-ray emission from the NS photosphere (i.e. the X-ray burst), and a fraction of the energy may be spent by ejecting the NS envelope in a wind during the super-Eddington phase (e.g., \citealt{PP86}).

The four decades of observations have shown that most X-ray burst energy spectra resemble black bodies \citep{LvPT93}.
Correspondingly, NS atmosphere models predict spectral shapes that are well described by a diluted black body (e.g., \citealt{LTH86,SPW12}), as long as the atmospheres are hot and/or metal-poor \citep{NSK15}.
Since the earliest observations in the 1970s (e.g., \citealt{HCL80}), for some X-ray bursts the black body model fails to fit the spectra in certain sections of the burst (e.g., \citealt{vPDT90, KHvdK02, iZW10}).
While some of these discrepancies can also originate from variations of the persistent/accretion spectrum during the bursts \citep{WGP13}, the statistical analysis by \citet{WGP15} suggests that there remains a considerable fraction of bursts where more complex spectral descriptions are needed.
By adding photo-ionization edges to the black body model one can achieve acceptable fits, but it is often difficult to be certain if these features are formed in the NS photosphere \citep{iZW10} or by reflecting the X-ray burst emission from the surrounding accretion disc \citep{SB02,KBK14}.

One of the peculiar bursts where the black body model clearly does not fit the data was observed from \hete\ by the \textit{Rossi X-ray Timing Explorer} (\textit{RXTE}) on 2005 July 21.
\citet{GMC08} showed that the spectral fits can be improved by including Comptonization and a variable local absorption column in the model.
Here we show that the spectra are naturally explained if there is a high metal fraction at the photosphere, consistent with the hypothesis of \citet{WBS06}, who discussed how nuclear burning ashes can rise up towards the photosphere by strong convection, that is also seen in recent multi-dimensional hydro-dynamical simulations \citep{MNA11,MZN14,ZMN15}. 
At the same time the outer H/He-rich NS envelope can be ejected, ultimately revealing the burning ashes.
The high iron abundance in the ashes is a likely reason for appearance of the edges in the spectra (see also \citealt{iZW10}) and, additionally, the metals change the photospheric opacity \citep{NSK15} that leads to the peculiar variations in the black body radius and temperature.

\section{Target, observations and spectral models}

\hete\ was discovered as a new X-ray transient on 2005 June 14 by the \textit{HETE-2} spacecraft, and follow-up \textit{RXTE} observations revealed it to be a 377~Hz accreting millisecond pulsar in a $83.3$~min binary orbit \citep{KMV06, SKT07}.
The pulsations were seen only intermittently, and the pulse amplitudes tended to increase right after X-ray bursts were observed \citep{GMK07}.
The magnetic field may have been buried by the accretion flow, causing the pulsations to cease within a few months from the discovery \citep{Patruno12}.
X-ray burst oscillations close to the pulsar's spin frequency were seen only in one observation when the source was in the soft spectral state in 2009 \citep{WAL09}.
The optical counterpart is a low-mass star ($M_\textrm{c} \lesssim 0.1 \Msun$) with clear hydrogen emission lines, and the binary inclination is $i \lesssim 20\degr$ \citep{ECF08}.
The distance to the source $d$ lies in the range 3.0-5.5 kpc \citep{GMH08}, and we assume here $d=4.2$ kpc, in the middle of the proposed range.

In this letter we report our analysis of the 10 X-ray bursts of \hete\ that have been observed by the \xte\ satellite (see \citealt{KNL14} table A1), mainly concentrating on the most energetic 2005 July 21 burst (Burst 1 hereafter).
We performed time-resolved X-ray spectral modeling of the Proportional Counter Array (PCA; \citealt{JMR06}) data in the 3--20~keV energy range.
The observations were split into 0.25~s to 2~s time segments. 
Initially the net burst spectra were fitted in {\sc xspec} \citep{Arnaud1996} with a black body model {\sc bbodyrad}, characterized by a temperature $T_\textrm{bb}$ and a normalization $K = (\Rbb\,[\textrm{km}] /d_{10})^2$, where $\Rbb$ is the black body radius and $d_{10}$ is the distance in units of 10~kpc.
Absorption by the interstellar medium was taken into account using the {\sc tbabs} model (with fixed hydrogen column density $N_\textrm{H} = 0.16\times10^{22}\,\mathrm{cm^{-2}}$, \citealt{WGP13}).
These data were also analyzed in \citet{KNL14}, where the standard data reduction methods are described.

\begin{figure}
\includegraphics[width=\linewidth]{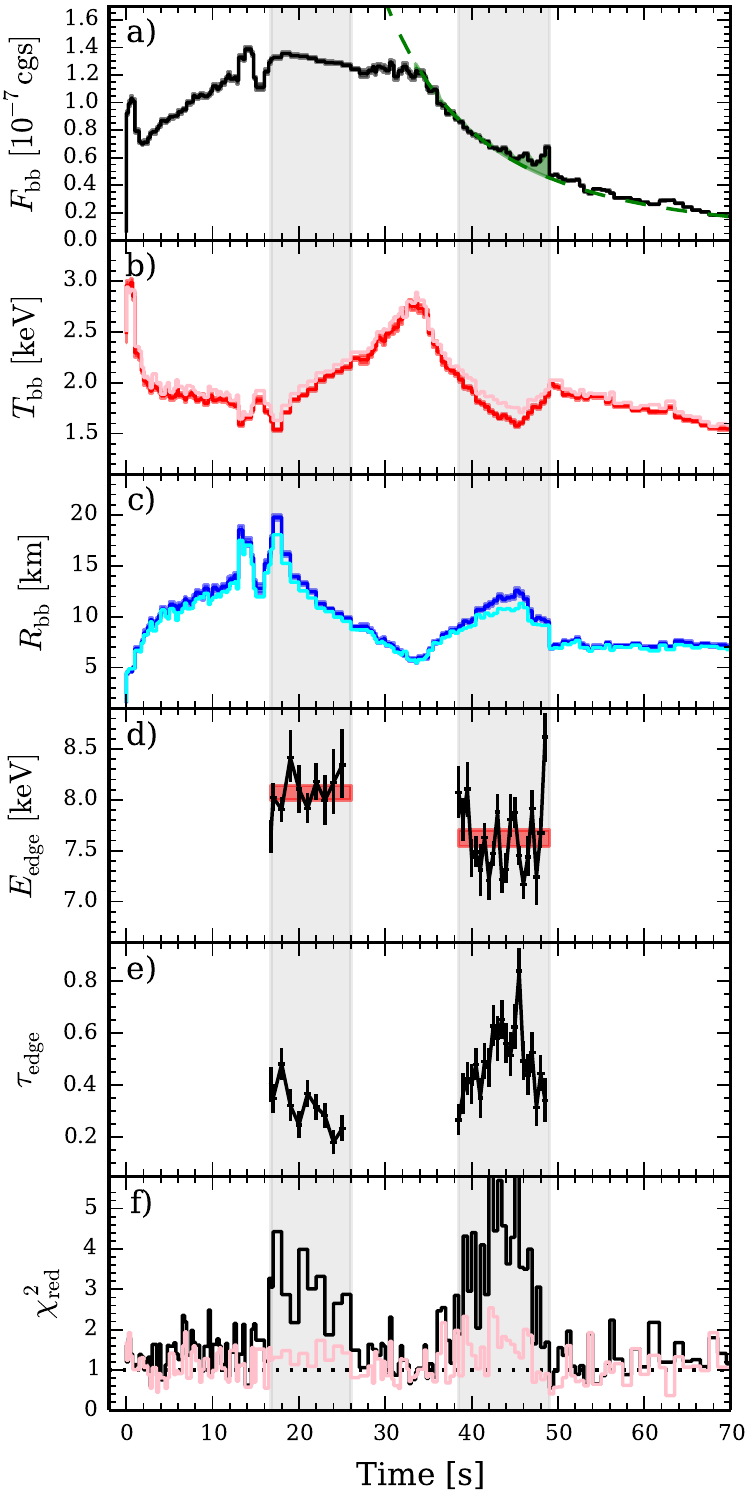}
\caption{\label{fig:bbparams} Time evolution of the best fitting spectral parameters in Burst 1. a) black body flux (the green line shows a power-law flux decay fit), b) black body temperature (red for {\sc bb} and pink for {\sc bb $+$ edge} model), c) black body radius assuming $d=4.2$ kpc (blue for {\sc bb} and cyan for {\sc bb $+$ edge} model), d) photo-ionization edge energy, e) edge optical depth and f) reduced $\chi^2$ of the fits (black for {\sc bb} model and pink for the {\sc bb $+$ edge} model). The width of the lines in panels a, b and c denote the 1$\sigma$ errors.}
\end{figure}

We additionally fitted the Burst 1 with our atmosphere model {\sc matmos} \citep{NSK15}.
The main model assumption is that the NS photosphere coincides with the NS surface, which is why we fit the model only in the cooling phase of the X-ray burst.
Stellar rotation is not considered here; this causes the photo-ionization edges to be (artificially) sharp features in the modeling, but the expected Doppler broadening is not larger than the spectral resolution. 
{\sc matmos} has two free parameters; the normalized atmospheric luminosity $g_{\rm rad}/g$ and the `metal enhancement factor' $\zeta$.
The first parameter is the ratio of the radiative acceleration $g_{\rm rad}$ and surface gravity $g$, where the value of unity corresponds to the Eddington limit. 
The second $\zeta$-parameter is defined as the factor by which metal abundances of \citet{AGS09} are increased, after which hydrogen and helium abundances are scaled down in solar ratios.
For the solar composition $\zeta \equiv 1$, and for an atmosphere consisting of metals only $\zeta \approx 70$.
The {\sc matmos} model normalization is $A=(R_{\infty}/d)^2$, where $R_{\infty} = \Rns (1+z)$ is the apparent NS radius as seen by a distant observer and $1+z = (1-2G\Mns/\Rns c^2)^{-1/2}$ is the gravitational redshift at the NS surface. 
We implicitly assume a NS radius of $\Rns = 12~\mathrm{km}$ and a mass of $\Mns = 1.4~\mathrm{M}_{\odot}$ and, therefore, with the constant distance of 4.2 kpc $A$ is not allowed to vary in the modeling.
Our choice of NS radius and mass then also determines the gravitational redshifts of the photo-absorption edges, which we find are well fit with this choice of NS mass and radius.
A more detailed analysis exploring the constraints on mass, radius and distance will be presented in a forthcoming paper.
{\sc matmos} was not imported into {\sc xspec}, but instead the model parameter minimization was performed in a Bayesian framework using the Metropolis-Hastings algorithm within the Markov Chain Monte Carlo method.

\section{Spectral analysis of the X-ray bursts}

The parameter time evolution resulting from the black body spectral modeling is shown in Fig. \ref{fig:bbparams} for the first and most energetic X-ray burst from HETE J1900.1--2455 (Burst 1), observed on 2005 July 21 (MJD 53572).
As noted by \cite{GMC08}, the light curve of this X-ray burst is highly atypical.
The burst rise-time from the persistent flux level $F_\textrm{per} \approx 0.7 \times 10^{-9}$ to $F_\textrm{bb} \approx 10^{-7}\,\ergcms$ was only about 30~ms, which is then followed by a 30 per cent flux drop in the next 2~s.
The subsequent increase of the flux (black line in Fig. \ref{fig:bbparams}a) is accompanied by a decreasing black body temperature (red line) and an increasing black body radius (blue line), indicating that the photosphere lifts off from the NS surface.
The radius expansion halts about 15~s into the burst, where the $F_\textrm{bb}$ and $R_\textrm{bb}$ attain two local maxima, while $T_\textrm{bb}$ simultaneously attains two local minima.
During the next 18~s the photosphere slowly contracts back towards the NS surface, finally reaching it at 34~s from the start of the burst.
This marks the photospheric `touchdown', where $T_\textrm{bb}$ reaches a local maximum and $R_\textrm{bb}$ a local minimum.
At the same time the flux decays from $F_\textrm{PRE} \approx 1.36 \times 10^{-7}$ to $F_\textrm{TD} \approx 1.23 \times 10^{-7}\,\ergcms$ during this photospheric radius expansion (PRE) phase.
The cooling phase of the burst has two very distinct phases. 
The initial cooling phase from 34~s onward shows a smooth flux decay with a gradual increase of the radius. 
Then around 48~s the temperature increases significantly while simultaneously $R_\textrm{bb}$ abruptly drops within 3~s.
This transition is accompanied by clear flux fluctuations on top of the power-law type decay (with an index of $\gamma = 1.74 \pm 0.02$, see Fig. \ref{fig:bbparams}a, green dashed line).
Afterwards $R_\textrm{bb}$ remains constant while the temperature evolves more slowly than in the initial phase.

\begin{figure}
\includegraphics[width=\linewidth]{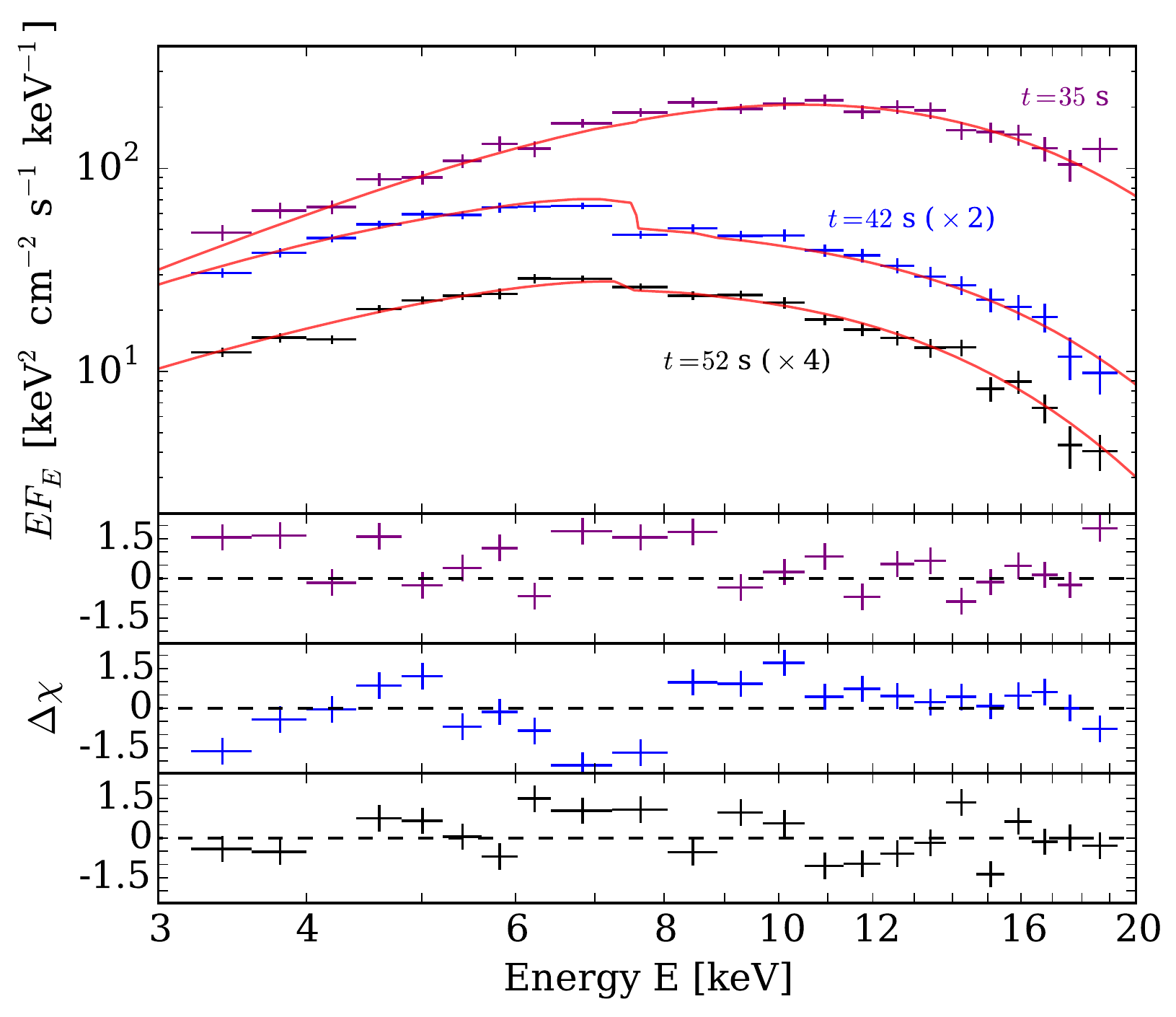}
\caption{\label{fig:spectra} $EF_\textrm{E}$ spectra of \hete\ X-ray burst 1 modeled with the {\sc matmos} atmosphere in 3 phases in the cooling tail.}
\end{figure}

The black body model is a poor description of the data in two localized parts in the radius expansion- and the cooling phase (Fig. \ref{fig:bbparams}f).
Similarly to \cite{iZW10}, we find that by adding a photo-absorption edge to the model (\textsc{edge} in \textsc{xspec}), the fits are improved significantly (Fig. \ref{fig:bbparams}def).
The addition of an \textsc{edge} does not change the black body temperature or radius significantly (pink and magenta lines in Fig. \ref{fig:bbparams}bc).
For this burst the edges appear significant only when the black body temperature is below about 2.2~keV (gray bands in Fig. \ref{fig:bbparams}).
Furthermore, the edge optical depth varies in the range of $\tau_\textrm{edge} \approx 0.2 - 0.6$ and it is anti-correlated with the black body temperature.
We also find that the edge energies are higher in the radius expansion phase as compared to the cooling phase, $8.07 \pm 0.08$ and $7.62 \pm 0.09$~keV, respectively.
The variable edge energy and depth, and their direct dependency upon the black body temperature (and not with the ionizing flux), strongly imply that the edges are produced in the stellar photosphere, rather than by reflecting the burst emission from the surrounding accretion disc.
The likely cause for the temperature dependency is that at higher temperatures (around touchdown) iron is fully ionized in the NS photosphere, and only at lower temperatures iron can recombine to create photo-ionization edges to the spectra \citep{NSK15}.

\begin{figure}
\includegraphics[width=\linewidth]{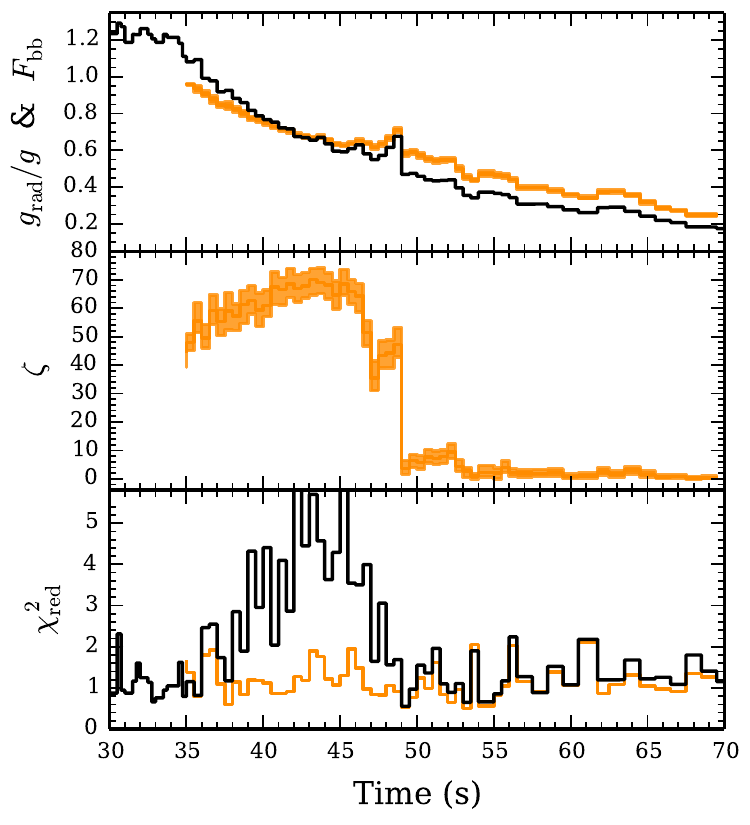}
\caption{\label{fig:matmosparams} {\sc matmos} model fits for the cooling phases with a variable chemical composition (orange lines). 
Top panel shows the normalized atmospheric luminosity, $g_{\rm rad}/g$, that decays similarly to the black body flux (black lines in units of $10^{-7}\,\ergcms$). 
The middle panel shows metal enhancement factor $\zeta$. 
The bottom panel shows the reduced $\chi^2$ for the {\sc matmos} and {\sc bbodyrad} models.}
\end{figure}

To study the edge features we fitted the cooling phase spectra with the atmosphere model {\sc matmos} (see Figs \ref{fig:spectra} and \ref{fig:matmosparams}).
The normalized atmospheric luminosity, $g_{\rm rad}/g$ in the {\sc matmos} model, decreases in a similar fashion as the black body flux, displaying the same small flux fluctuations towards the end of the cooling phase.
The most remarkable feature is the metal enhancement factor, which is initially higher than about $\zeta \approx 50$ and it is consistent with the maximum value of $\zeta \approx 70$ where the atmosphere consists entirely of metals (i.e. no hydrogen nor helium).
Then, around 48~s after the burst onset, the metal content drops towards the solar value $\zeta \approx 1$ in about 3~s.
The {\sc matmos} model gives a far better description of the data, $\chi^2_{\rm red} \approx 1.2$, compared to $\chi^2_{\rm red} \gtrsim 3$ in the black body fits with the same number of free parameters (see Fig. \ref{fig:matmosparams}).

The atmosphere modeling shows that the edge is mainly produced by H-like Fe ions at $E_\textrm{Fe \textsc{XXVI}} = 9.28$~keV, that is gravitationally redshifted to $7.62$~keV.
While only the H-like iron contributes significant edges in the 3--20~keV range, the other metals impact the spectra in a more indirect way.
The metals increase the photospheric electron density and the bound-free and free-free opacities, pushing the spectral color correction factor -- defined as the ratio between the colour- and the effective temperatures $\fc \equiv T_\textrm{c}/T_\textrm{eff}$ -- to lower values with respect to solar abundances \citep{NSK15}.
Therefore, the drop in the metal enhancement causes the color-correction factor to jump from $\fc\approx 1.1$ to $1.4$, which results in the observed black body temperature and radius variations seen in Fig. \ref{fig:bbparams} around 48~s, given the dependency $\Rbb = \Rns (1+z)\fc^{-2}$ \citep{LvPT93}.

\begin{figure}
\includegraphics[width=\linewidth]{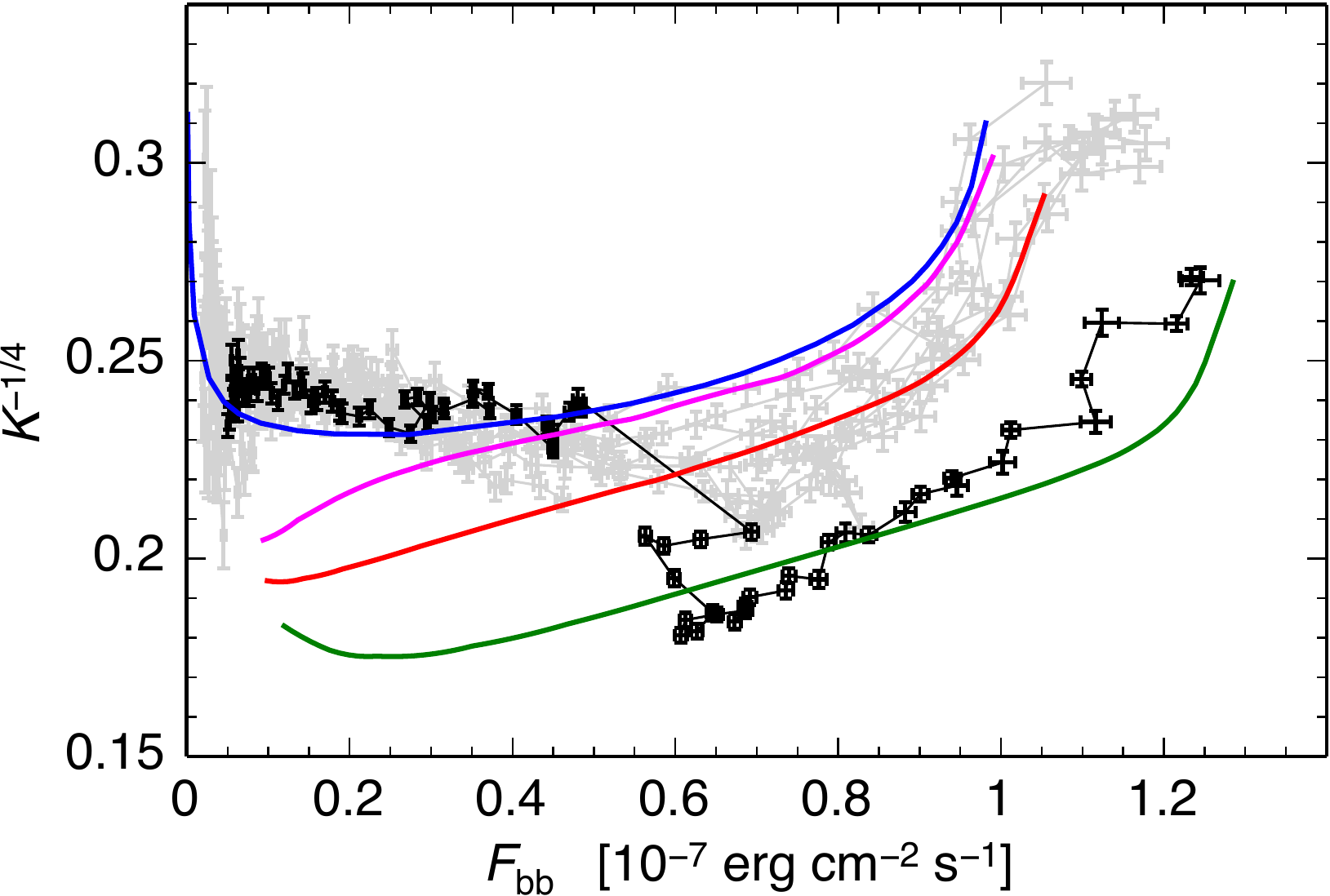} 
\caption{\label{fig:fcs} 
Black body normalization versus the flux of \hete\ bursts observed with the PCA instrument, together 
with the theoretical color correction curves, $\fc$, for a $\Mns = 1.4\,\Msun$ and $\Rns = 12$~km NS.
The track from the 2005 July 21 burst is high-lighted with black color. The blue, purple, red and green lines are for {\sc matmos} $\fc$ curves with metal enhancement factors $\zeta = 0.01$, $1$, $10$ and $40$, respectively.}
\end{figure}

Other X-ray bursts have been observed from \hete, and so they also could have exhibited similar metal-induced variations in $\fc$, that would be imprinted as a scatter in the observed black body radii.
For this reason we compare the emission areas during the cooling tails of all hard state X-ray bursts from \hete\ in Fig. \ref{fig:fcs}, in the so-called $l$--$\fc$ diagram (equivalently $F$--$K^{-1/4}$ diagram).
This is a convenient way to compare many X-ray bursts to atmosphere models, because of the relation $\fc \propto K^{-1/4}$ \citep{LvPT93,SPR11}.
In Fig. \ref{fig:fcs}, Burst 1 is highlighted with black symbols, while the other 7 bursts which occurred in the hard state are shown with gray symbols.
The discrepant evolution of the soft- and intermediate state bursts -- likely due to the spreading layer in the disc-NS boundary \citep{SPR11, PNK14, KKK16} -- is not shown in Fig. \ref{fig:fcs}; such bursts start with a constant normalization, and then exhibit a large drop at lower fluxes, similarly to the soft state bursts of 4U 1636--536 \citep{ZMA11} and 4U 1608--52 \citep{PNK14}.
The blue line shows the theoretical $\zeta=0.01$ $\fc$-curve, which is in good agreement with the data for the bursts at fluxes below about 50 per cent of the Eddington value.
At higher fluxes, the bursts gradually deviate from the low metallicity $\fc$-curves, each becoming more consistent with moderately enhanced atmospheres ($\zeta=[1-10]$; purple and red lines).
Burst 1, clearly standing out as the most extreme case, is initially closer to the $\zeta=40$ curve and then jumps up to values consistent with solar to sub-solar tracks during the transition in the cooling phase.
Interestingly, most of the bursts are consistent with metal enhanced ($\zeta=[1-10]$) $\fc$-curves closer to the touchdown flux, and they all tend to follow the sub-solar curves below $F_\textrm{bb} \lesssim 4 \times 10^{-8}\,\ergcms$.

\section{Discussion}

The appearance of edges $14$~s after the burst onset likely marks the time when the outer H/He-rich envelope is ejected in the wind \citep{WBS06}. Based on the light curve models of \citet{iZW10} (see their Fig.~C1), the fact that the bursts stayed above Eddington luminosity for more than 30 seconds implies an ignition column $y\approx 2 \times 10^9 \, \gcm2$. 
This can be accumulated in about 16 days with the persistent accretion rate observed prior to the X-ray burst (roughly 2 per cent of Eddington), matching well with the HETE-2 detection of a previous X-ray burst 14 days earlier \citep{SKT07}. 
At these low accretion rates stable hydrogen burning likely built up a large He pile \citep{GC06}, consistent with the strong radius expansion.
The expected wind mass loss rate is $\dot{M}_\textrm{w}\sim 10^{18}\,\gs$ \citep{PP86,WBS06} which in 14 seconds can eject a mass $M_\textrm{w} > 10^{19}$~g. 
This matches the column of $y_\textrm{w} \approx 10^6\ {\rm g\ cm^{-2}}$ that must be ejected to expose ashes, set by the maximum extent of the convection zone during the flash \citep{WBS06}. 
Thus, the deep ignition and the unusually long PRE-phase likely provided the necessary conditions for the convection to extend sufficiently far out towards the photosphere.
In the subsequent 18~s `metal-enriched wind phase' the total mass loss is also $\Delta M_{\rm z} \sim 10^{19}$~g, which, according to the \textsc{matmos} model results, consists largely of metals.
The large He fraction before ignition \citep{FPB07} means a low proton abundance which suggests that significant rp-process burning did not occur in this burst \citep{WBS06}.
Therefore, it remains unclear if the rp-process contributes to the Galactic metal enrichment as suggested by \citet{SAG98}.

When the temperature was below about $\lesssim 2.2$~keV, photo-absorption edges appeared both in the Eddington-limited phase and in the cooling phase, and the edge energies decreased from $8.07 \pm 0.08$ to $7.62 \pm 0.09$~keV, respectively.
Our atmosphere models showed that in the cooling phase the dominant ion that produces the photo-ionization edge is H-like Fe at $E_{\textrm{Fe} \textsc{xxvi}} = 9.28$~keV, although other weaker features are also present.
This energy shift $E_{\textrm{Fe}\,\textsc{xxvi}} / E_\textrm{edge} \approx 1.22$ is consistent with the gravitational redshift at the NS surface, $1+z \approx 1.24$, for the assumed $\Mns=1.4\,\Msun$ and $\Rns=12$~km.
In the Eddington limited phase the redshift is instead $1+z_\textrm{PRE} \approx 1.15$.
The measured redshift difference can then be used to estimate how much the photosphere expanded during the Eddington-limited phase.
We find that the radius where the edge is formed is about $R_\textrm{PRE} \approx 17$~km, i.e. the radius expansion was roughly 5~km. 
Note also that a similar redshift variation can be derived from the relation $F_\textrm{PRE} (1+z_\textrm{PRE})^2 = F_\textrm{TD} (1+z_\textrm{TD})^2$ (e.g., \citealt{LvPT93}), which assumes that in the radius expansion phase the flux is always the local Eddington flux.
Furthermore, given that $\Rbb = \Rns (1+z)\fc^{-2}$ \citep{LvPT93}, we can also estimate the color correction factor during the radius expansion phase.
At touchdown, we have measured $1+z_\textrm{TD} \approx 1.22$, $R_\textrm{bb,TD} \approx 5.5$~km and taking $\Rns = 12$~km we obtain $f_\textrm{c,TD} \approx 1.63$, which is consistent with highly metal enriched atmosphere models near the Eddington flux \citep{NSK15}.
In the radius expansion phase, we can use the estimated $1+z_\textrm{PRE} \approx 1.15$, $R_\textrm{bb,PRE} \approx 16.3$~km (at 19~s after the burst onset) and $R_\textrm{PRE} \approx 17$~km to find that $f_\textrm{c,PRE} \approx 1.10$.
The ratio $f_\textrm{c,PRE} / f_\textrm{c,TD} \approx 0.67$ (independent of distance), is similar to the values found in the cooling tail before the transition at about 50 per cent of the Eddington flux. 
The presence of strong photo-ionization edges in cooler photosphere in both cases likely reduces the color correction factor to the same value, but accurate radiative transfer calculations for expanding, wind launching NS atmospheres are needed to confirm this behavior.

In Fig. \ref{fig:bbparams}a the flux enhancements between 46--49~s after the burst onset have an excess fluence of about $3.9 \times 10^{-8}~\mathrm{erg}\,\mathrm{cm}^{-2}$ above the power-law decay (indicated by the green hatched area in Fig. \ref{fig:bbparams}).
The strong radius expansion likely halted accretion onto the NS before this time, and the ``missing'' persistent emission fluence accumulated would have been about $3.3\times 10^{-8}~\mathrm{erg}\,\mathrm{cm}^{-2}$. 
The match between these fluences suggests that the flux excess in the tail is caused by accreting the accumulated gas that was halted by the X-ray burst.
Therefore, the transition to the second cooling phase in the X-ray burst was likely caused by covering the metal-rich NS photosphere with fresh, H/He-rich layer from the accretion disc.
The $\sim$10~s duration of the metal-rich cooling stage allowed the photosphere to cool enough to reveal the ashes as strong photo-absorption edges.
This was likely facilitated by the unusually long PRE-phase, during which the accretion disc was pushed a few kilometers backwards by the X-ray burst.
In other bursters (and in the other \hete\ bursts) with shorter PRE-phases and/or weaker magnetic fields, the NSs may be covered by a H/He-rich layer before the photosphere cools enough, and thus they may not have as obvious non-Planckian spectra.

The detection of burning ashes in \hete, and likely in other bursters \citep{iZW10}, has several consequences.
We have found evidence that the burning ashes should get convectively mixed in the NS envelope during X-ray bursts (as simulations suggest), and that NSs should have significant mass loss during radius-expansion episodes.
More detailed modelling will involve coupling burst models which follow convection and nuclear burning (e.g., \citealt{WHC04}) to models of super-Eddington winds \citep{PP86}. 
Also, input from nuclear experiments is important to include in nuclear networks to make accurate predictions of the ash composition \citep{WBS06}.
Another consequence is related to the observed $\Rbb$ scatter and the touchdown flux scatter seen in several bursters \citep{GPO12,GOP12}, which can be caused by the variable metal enhancement in different X-ray bursts.
At least for \hete, judging from Fig. \ref{fig:fcs}, most of the PRE-bursts show signs of photospheric metal enhancement, and thus the nuclear burning ashes may reach the NS photospheres much more commonly than previously thought.
If this turns out to be the case, then our results suggest that the use of black bodies and color-correction factors in deriving NS $M$--$R$ constraints from X-ray burst observations must be done with care: it is not obvious that the typical X-ray burst is one where no metal enhancement is present (compare, e.g., \citealt{SPR11,NSK16}).
The ashes may lead to systematic uncertainties in NS radii by up to 4~km (see \citealt{PNK14}, fig. 7).
On the other hand the photo-absorption edges, where present, provide an independent $M$--$R$ constraint, but that requires fitting the atmosphere models directly to the observed spectra. 
For better measurements of edges for bursts with more modest metal enhancements (and of other elements than iron), an observatory like the \textit{LOFT} (see, e.g., \citealt{iZAB15}) would be needed.
Meanwhile, analysis of archival data sets should give us a better handle on how frequently, and under which conditions, the edges can be detected.

\section*{Acknowledgements}

We thank the referee for a careful review that helped to improve the manuscript.
JJEK acknowledges support from the ESA research fellowship programme.
JN acknowledges support from the University of Turku Graduate School in Physical and Chemical Sciences, and the support from the Faculty of the European Space Astronomy Centre (ESAC).
JP was supported by the Foundations' Professor Pool, the Finnish Cultural Foundation and the Academy of Finland grant 268740.
AC is supported by an NSERC Discovery grant and is an Associate Member of the CIFAR Cosmology and Gravity program.
VS was supported by the German Research Foundation (DFG), grant WE 1312/48-1 and the Russian Foundation for Basic Research (grant 16-02-01145-a).
This research was undertaken on Finnish Grid Infrastructure (FGI) resources.
We also acknowledge the support of the International Space Science Institute (Bern, Switzerland) and ``NewCompStar'' COST Action MP1304.




\bibliographystyle{mnras}

\bsp	
\label{lastpage}
\end{document}